\documentclass[12pt]{article}
\usepackage[textures]{graphics}
\parskip=\bigskipamount
\begin{document}
\title{\bf Quantum entanglement, interaction, and the classical limit.}
\date{}
\author{}
\maketitle
\vglue -1.8truecm
\centerline{\large Thomas Durt\footnote{TENA-TONA Free University of
Brussels, Pleinlaan 2, B-1050 Brussels, Belgium. email: thomdurt@vub.ac.be}}\bigskip\bigskip


{\it Abstract: Two or more quantum systems are said to be in an entangled or non-factorisable
state if their joint (supposedly pure) wave-function is not expressible as a product of individual
wave functions but is instead a superposition of product states. It is only when the systems are
in a factorisable state that they can be considered to be separated (in the sense of Bell).
We show that whenever two quantum systems interact with each other, it is impossible that all
factorisable states remain factorisable during the interaction unless the full Hamiltonian does
not couple these systems so to say unless they do not really interact. We also present certain
conditions under which particular factorisable states remain factorisable although they represent a
bipartite system whose components mutually interact. We identify certain quasi-classical regimes that satisfy 
these conditions and show that they correspond to classical, pre-quantum, paradigms associated to the concept of particle}

\section*{Introduction}

The term entanglement was first
introduced by Schr\"odinger who described this as the characteristic trait of quantum mechanics,
``the one that enforces its entire departure from classical
lines of thought'' \cite{1}. Bell's inequalities \cite{2} show that when two systems are prepared in an entangled
state, the knowledge of the whole cannot be reduced to the knowledge of the parts, and that to
some extent the systems lose their individuality. It is only when their joint wave-function is
factorisable that they are separable\footnote{It can be shown that whenever two distant systems are
in an entangled (pure) state, there exist well-chosen observables such that the associated
correlations do not admit a local realist explanation, which is revealed by the violation of
well-chosen Bell's inequalities \cite{3,4}.}. It is therefore interesting to investigate which are the
situations such that two systems, initially prepared in a (pure) product state remain in such a state although
they mutually interact. We shall show that when the Hilbert spaces associated
to the interacting systems
$A$ and
$B$ are finite dimensional, if we impose that all the product states remain product states during
the interaction, the full Hamiltonian can be factorised as follows:
$\label{fact}H_{AB}(t)=H_{A}(t)\otimes I_{B}+I_{A}\otimes H_{B}(t)$, where
$H_{i}(t)$ acts on the ``$i$'' system only while $I_{j}$ is the identity operator on the ``$j$''
system ($i,j=A,B$). In other words, in quantum mechanics there is no interaction without
entanglement. We shall also present a sufficient condition under which particular factorisable
(non-necessary pure) states remain factorisable during the interaction. Finally we shall show that
 when two particles obey the non-relativistic Schr\"odinger equation, we can distinguish three regimes
  for which this condition is satisfied, that correspond to the following classical paradigms:  material point,
   test-particle, and diluted particle (droplet model). We reinterpret these results at the light of the ''predictability sieve'' criterion (PS criterion) proposed by Zurek 
   in the decoherence approach \cite{Zur1,Zur2, Zur3}
    and note that, although they are obtained in an oversimplified quantum modellisation of a system and an environment 
   (here, the universe consists of two particles considered in first quantisation that interact through a position-dependent
    potential in the non-relavistic regime), these results confirm the basic intuition of the PS criterion:
     the classical behavior corresponds to the islands of the Hilbert space characterized by the minimal Shannon-von Neumann entropy
      (or equivalently by the maximal coherence of the reduced state of the system obtained after tracing out the environment).

\section{Two interacting finite-dimensional systems: entanglement versus interation.}

Let us consider two interacting quantum systems $A$ and $B$. We assume that the Hilbert
spaces associated to these systems are finite dimensional (of dimensions $d_A$ and $d_B$
respectively), that the wave-function of the full system is a pure state of ${\bf
C}^{d_A}\otimes {\bf C}^{d_B}$ and obeys the Schr\"odinger equation: 
\begin{equation}\label{schrod}i \hbar\, \partial_t\, {\bf \Psi}_{AB}( t) =
H_{AB}( t){\bf \Psi}_{AB}( t)\end{equation}where $H_{AB}( t)$ is a self-adjoint
 operator that acts on ${\bf C}^{d_A}\otimes {\bf C}^{d_B}$, that we assume to be sufficiently regular in time in order
to ensure
 that the temporal Taylor development of the
wave-function is valid up to the second order in time.

{\bf Main Theorem:}

All the product states
remain product states during the interaction if and only if the full Hamiltonian can be
factorised as follows:
\begin{equation}\label{fact}H_{AB}(t)=H_{A}(t)\otimes I_{B}+I_{A}\otimes H_{B}(t)\end{equation}where
$H_{i}$ acts on the $i$th system only while $I_{j}$ is the identity operator on the $j$th system
($i,j=A,B$).

In order to prove this theorem, we shall firstly prove the following lemma:

{\bf Lemma:}

A pure product state
remains product state during the interaction if and only if, during its evolution, the
Hamiltonian never couples this product state to a product state that is bi-orthogonal to it.

{\bf Proof of the Lemma:}

A) Proof of the necessary condition. Let us assume that a state initially (at time $t$) factorisable
 remains a product state throughout the evolution:
  ${\bf \Psi}_{AB}( t') =
\psi_{A}( t')\otimes\psi_{B}( t'), \forall t' \geq t$. Then, in virtue of
the Leibnitz rule for the time derivative of a product, we get that $H_{AB}(t'){\bf \Psi}_{AB}( t')=
{1\over i\hbar }\partial_{t'}{\bf \Psi}_{AB}( t') =
{1\over i\hbar }(\partial_{t'}(\psi_{A}( t'))\otimes\psi_{B}( t')+\psi_{A}( t')\otimes \partial_{t'}(\psi_{B}( t')))$. The former state 
is obviously not biorthogonal to $\psi_{A}( t')\otimes\psi_{B}( t')={\bf \Psi}_{AB}( t')$; actually it belongs to the space which is orthogonal to 
the space that contains the states biorthogonal to ${\bf \Psi}_{AB}( t')$. In appendix, we show that when a state is factorisable, 
the squared lenght of the component of $H_{AB}(t'){\bf \Psi}_{AB}( t')$ that is biorthogonal to this state is proportional 
to the rate of decrease of the trace of  the squared reduced density matrix, where the reduced density matrix is obtained
from the full density matrix after tracing out
 one of the subsystems ($A$ or $B$). This rate is also equal to the linear entropy production, where the linear entropy
  of the reduced state $\rho$ is defined as $Tr\rho-Tr\rho^2=1-Tr\rho^2$. It is a good measure of the degree of 
  entanglement between $A$ and $B$, when the full state is pure, and nearly factorisable (it coincides with the linear term in
the Taylor development of the Shannon-von Neumann entropy of the reduced density matrix, the constant term being equal
to zero for factorisable states). Formulated so, the physical meaning of the necessary condition is very trivial: 
if the
full state remains factorisable throughout time, then no
entanglement occurs (the rate of creation of the entanglement is equal to zero).

B) Proof of the sufficient condition. Let us consider that at time $t$ the system is
prepared in a product state ${\bf \Psi}_{AB}( t) =
\psi_{A}( t)\otimes\psi_{B}( t)$. Let us choose a
basis of product states $\big|\psi^i_{A}\big>\otimes \big|\phi^j_{B}\big>$ similar to the basis
introduced in the appendix, so to say a
basis of product states $\big|\psi^i_{A}\big>\otimes \big|\phi^j_{B}\big>$ ($i: 1...d_A;
j:1...d_B$, and $\big<\psi^i_{A}\big|\psi^{i'}_{A}\big> =
\delta_{ii'}; \big<\phi^j_{B}\big|\phi^{j'}_{B}\big>=\delta_{jj'})$  such that $\psi_{A}( t) = \big|\psi^1_{A}\big>$
and
$\phi_{B}( t) = \big|\phi^1_{A}\big>$. Let us define the Hamiltonian by its matricial elements 
n this basis as follows: 
\begin{equation}
\label{diantre} H_{ikjl}=
\big<\psi^i_{A}\big|\otimes\big<\phi^j_{B}\big|H_{AB}(t)\big|\psi^k_{A}\big>\otimes\big|\phi^l_{B}\big>+
\tau(\delta t^2)\end{equation}When the Hamiltonian does not couple ${\bf
\Psi}_{AB}( t)$ to states that are bi-orthogonal to it,
$\Sigma_{i: 2...d_A;j:2...d_B}\big|H_{i1j1}\big|^2= 0$ (where $H_{ikjl}$ is defined in eqn.
\ref{diantre}) and, in virtue of eqn. \ref{equation}:
\begin{equation}i \hbar\, \partial_t\, {\bf \Psi}_{AB}( t) =
H_{AB}( t){\bf \Psi}_{AB}( t)= (\Sigma_{i:
1...d_A}H_{i111}\big|\psi^i_{A}\big>)\otimes
\big|\phi^1_{B}\big>+\big|\psi^1_{A}\big>\otimes(\Sigma_{j:
2...d_B}H_{11j1}\big|\phi^j_{B}\big>)\end{equation} We can rewrite this equation as follows:
\begin{equation}\label{eff3}i \hbar\, \partial_t\, {\bf \Psi}_{AB}( t) = (H^{eff.}_A( t)\cdot\psi_{A}(
t))\otimes\psi_{B}( t)+\psi_{A}( t)\otimes(H^{eff.}_B( t)\cdot \psi_{B}( t))\end{equation} 
where the effective Hamiltonians $H^{eff.}$ are defined as follows:
\begin{equation}\label{eff1}H^{eff.}_A(t)\cdot\rho_{A}( t)=Tr_B( H_{AB}(
t){\bf
\rho}_{AB}( t))\end{equation} and
\begin{equation}\label{eff2}H^{eff.}_B(t)\cdot
\rho_{B}( t)=Tr_A (H_{AB}( t){\bf \rho}_{AB}( t))-(Tr_{AB} (H_{AB}( t){\bf \rho}_{AB}(
t)))\cdot\rho_{B}( t)\end{equation} In these expressions $Tr_{i}$ represents the partial trace over
the degrees of freedom assigned to the system $i$ while 
${\bf
\rho}_{AB}( t)$ is the projector onto
${\bf
\Psi}_{AB}( t)$, $\rho_A(t)=Tr_B{\bf
\rho}_{AB}( t)$, and $\rho_B(t)=Tr_A{\bf
\rho}_{AB}( t)$. For instance, we have that \begin{displaymath}Tr_B( H_{AB}(
t){\bf
\rho}_{AB}( t))= \Sigma_{l: 1...d_B}\big<\phi^l_{B}\big| H_{AB}\big|\psi^1_{A}\big>\otimes
\big|\phi^1_{B}\big> \big<\psi^1_{A}\big|\otimes
\big<\phi^1_{B}\big| \phi^l_{B}\big>\end{displaymath}
\begin{displaymath} = \Sigma_{l:
1...d_B}\big<\phi^l_{B}\big| \Sigma_{i: 1...d_A,j:
1...d_B}H_{i1j1}\big|\psi^i_{A}\big>\otimes\big|\phi^j_{B}\big>
\delta_{l1}\big<\psi^1_{A}\big|\end{displaymath}
\begin{displaymath} = \Sigma_{l:
1...d_B} \Sigma_{i: 1...d_A,j:
1...d_B}H_{i1j1}\big|\psi^i_{A}\big>\delta_{lj}
\delta_{l1}\big<\psi^1_{A}\big|=(\Sigma_{i:
1...d_A}H_{i111}\big|\psi^i_{A}\big>\big<\psi^1_{A}\big|)\end{displaymath}
 so that $H^{eff.}_A( t)\cdot\psi_{A}(
t)=\Sigma_{i:
1...d_A}H_{i111}\big|\psi^i_{A}\big>$.
Similarly, we get that $H^{eff.}_B( t)\cdot\psi_{B}(
t)=\Sigma_{i:
2...d_B}H_{1i11}\big|\psi^i_{B}\big>$

Let us consider the product state $\psi_{A}^{red}( t')\otimes\psi_{B}^{red}( t')$, where
$\psi^{red}_{A(B)}( t')$ is a solution of the reduced Schr\"odinger equation
$i
\hbar\,
\partial_{t'}\, 
\psi^{red}_{A(B)}( t') =
 H^{eff.}_{A(B)}( t')\cdot\psi^{red}_{A(B)}( t')$ for the initial condition $\psi^{red}_{A(B)}( t)$
=
$\psi_{A(B)}( t)$. Obviously, $i \hbar\, \partial_{t'}\,
\psi_{A}^{red}( t')\otimes\psi_{B}^{red}( t')=H_{AB}\psi_{A}^{red}( t')\otimes\psi_{B}^{red}( t')$
and
${\bf
\Psi}_{AB}( t) =
\psi_{A}^{red}( t)\otimes\psi_{B}^{red}( t)$ so that, in virtue of the deterministic character of Schr\"odinger's
equation, ${\bf
\Psi}_{AB}( t') =
\psi_{A}^{red}( t')\otimes\psi_{B}^{red}( t')$, $\forall t' \geq t$ which ends the proof of the
lemma.

We shall now prove the main theorem.

{\bf Proof of the Main Theorem:}

A) Proof of the necessary condition. Let us choose a
basis of product states $\big|\psi^i_{A}\big>\otimes \big|\phi^j_{B}\big>$ ($i: 1...d_A;
j:1...d_B$ and $\big<\psi^i_{A}\big|\psi^j_{A}\big> =
\delta_{ij}=\big<\phi^i_{B}\big|\phi^j_{B}\big>)$.
If we impose that all the product states
remain product states during the interaction, then, in virtue of the lemma, the full
Hamiltonian never couples a product state to a product state that is bi-orthogonal to it. Then,
at any time $t$,
$\Sigma_{i: 2...d_A;j:2...d_B}\big|H_{i1j1}\big|^2= 0$ (where $H_{ikjl}$ is defined in eqn.
\ref{diantre}) so that we have that:
\begin{displaymath}H_{AB}(t)\cdot\big|\psi^i_{A}\big>\otimes
\big|\phi^j_{B}\big>=\big|\triangle_A^{ij}\psi^i_{A}\big>\otimes\big|\phi^j_{B}\big>+
\big|\psi^i_{A}\big>
\otimes\big|\triangle_B^{ij}\phi^j_{B}\big>\end{displaymath}where
\begin{equation}\big|\triangle_A^{ij}\psi^i_{A}\big>=\Sigma_{k:
1...d_A}H_{kijj}\big|\psi^k_{A}\big>\end{equation} and
\begin{equation}\label{deltaB}\big|\triangle_B^{ij}\phi^i_{B}\big>=\Sigma_{k: 1...d_B, k\not=
j}H_{iikj}\big|\phi^k_{B}\big>\end{equation}

Let us consider that at time $t$ the system is prepared along one of the first four states ${\bf
\Psi}^i_{AB}$ ($i:1,...4$) of this basis: ${\bf \Psi}^1_{AB}( t) =
\big|\psi^1_{A}\big>\otimes\big|\phi^1_{B}\big>$, ${\bf \Psi}^2_{AB}( t) =
\big|\psi^1_{A}\big>\otimes\big|\phi^2_{B}\big>$,
${\bf \Psi}^3_{AB}( t) = \big|\psi^2_{A}\big>\otimes\big|\phi^1_{B}\big>$, ${\bf \Psi}^4_{AB}(
t) =
\big|\psi^2_{A}\big>\otimes\big|\phi^2_{A}\big>$. Then, 
\begin{displaymath}H_{AB}(t)\cdot{\bf
\Psi}^1_{AB}(t)=\big|\triangle_A^{11}\psi^1_{A}\big>\otimes\big|\phi^1_{B}\big>+\big|\psi^1_{A}
\big>\otimes \big|\triangle_B^{11}\phi^1_{B}\big>\end{displaymath}
\begin{displaymath}H_{AB}(t)\cdot{\bf
\Psi}^2_{AB}(t)=\big|\triangle_A^{12}\psi^1_{A}\big>\otimes\big|\phi^2_{B}\big>+\big|\psi^1_{A}
\big>\otimes \big|\triangle_B^{12}\phi^2_{B}\big>\end{displaymath}
\begin{displaymath}H_{AB}(t)\cdot{\bf
\Psi}^3_{AB}(t)=\big|\triangle_A^{21}\psi^2_{A}\big>\otimes\big|\phi^1_{B}\big>+\big|\psi^2_{A}
\big>\otimes \big|\triangle_B^{21}\phi^1_{B}\big>\end{displaymath}
\begin{displaymath}H_{AB}(t)\cdot{\bf
\Psi}^4_{AB}(t)=\big|\triangle_A^{22}\psi^2_{A}\big>\otimes\big|\phi2_{B}\big>+\big|\psi^2_{A}
\big>\otimes \big|\triangle_B^{22}\phi^2_{B}\big>\end{displaymath}
By linearity, 
\begin{displaymath}H_{AB}(t)\cdot {1\over \sqrt 2}({\bf
\Psi}^1_{AB}(t)+{\bf
\Psi}^3_{AB}(t))=H_{AB}(t)\cdot {1\over \sqrt
2}(\big|\psi^1_{A}\big>+\big|\psi^2_{A}\big>)\otimes\big|\phi^1_{B}\big>\end{displaymath}\begin{displaymath}={1\over
\sqrt
2}((\big|\triangle_A^{11}\psi^1_{A}\big>+\big|\triangle_A^{21}\psi^2_{A}\big>)\otimes\big|\phi^1_{B}\big>+\big|\psi^1_{A}
\big>\otimes \big|\triangle_B^{11}\phi^1_{B}\big>+\big|\psi^2_{A}
\big>\otimes \big|\triangle_B^{21}\phi^1_{B}\big>)\end{displaymath}
\begin{displaymath}={1\over
\sqrt
2}((\big|\triangle_A^{11}\psi^1_{A}\big>+\big|\triangle_A^{21}\psi^2_{A}\big>)\otimes\big|\phi^1_{B}\big>
+(\big|\psi^1_{A}\big>+\big|\psi^2_{A}\big>)\otimes
(\big|\triangle_B^{11}\phi^1_{B}\big>+\big|\triangle_B^{21}\phi^1_{B}\big>)\end{displaymath}
\begin{displaymath}+(\big|\psi^1_{A}\big>-\big|\psi^2_{A}\big>)\otimes
(\big|\triangle_B^{11}\phi^1_{B}\big>-\big|\triangle_B^{21}\phi^1_{B}\big>)\end{displaymath}
${1\over \sqrt 2}(\big|\psi^1_{A}\big>-\big|\psi^2_{A}\big>)$
is orthogonal to ${1\over \sqrt 2}(\big|\psi^1_{A}\big>+\big|\psi^2_{A}\big>)$, so that
$H_{AB}(t)\cdot {1\over \sqrt 2}({\bf
\Psi}^1_{AB}(t)+{\bf
\Psi}^3_{AB}(t))$ couples ${1\over \sqrt 2}({\bf
\Psi}^1_{AB}(t)+{\bf
\Psi}^3_{AB}(t))$ to a bi-orthogonal state unless
$(\big|\triangle_B^{11}\phi^1_{B}\big>-\big|\triangle_B^{21}\phi^1_{B}\big>)$ is parallel to
$\big|\phi^1_{B}\big>$. Now, ${1\over \sqrt 2}({\bf
\Psi}^1_{AB}(t)+{\bf
\Psi}^3_{AB}(t))$ is a product state so that, in virtue of the lemma, the following constraint
must be satisfied:
\begin{displaymath}(\big|\triangle_B^{11}\phi^1_{B}\big>-\big|\triangle_B^{21}
\phi^1_{B}\big>)=\lambda\big|\phi^1_{B}\big>\end{displaymath}
The same reasoning is valid with the states ${1\over \sqrt 2}({\bf
\Psi}^2_{AB}(t)+{\bf
\Psi}^4_{AB}(t))$, ${1\over \sqrt 2}({\bf
\Psi}^1_{AB}(t)+{\bf
\Psi}^2_{AB}(t))$ and ${1\over \sqrt 2}({\bf
\Psi}^3_{AB}(t)+{\bf
\Psi}^4_{AB}(t))$ and leads to the following constraints:
\begin{displaymath}(\big|\triangle_B^{12}\phi^2_{B}\big>-\big|\triangle_B^{22}
\phi^2_{B}\big>)=\lambda'\big|\phi^2_{B}\big>\end{displaymath}

\begin{displaymath}(\big|\triangle_A^{11}\psi^1_{A}\big>-\big|\triangle_A^{12}
\psi^1_{A}\big>)=\lambda''\big|\psi^1_{A}\big>\end{displaymath}

\begin{displaymath}(\big|\triangle_A^{21}\psi^2_{A}\big>-\big|\triangle_A^{22}
\psi^2_{A}\big>)=\lambda'''\big|\psi^2_{A}\big>\end{displaymath}

By definition (eqn. \ref{deltaB}), $\big|\triangle_B^{ij}\phi^j_{B}\big>$ is orthogonal to
$\big|\phi^j_{B}\big>$ so that necessarily $\lambda=\lambda'=0$.
Let us now consider the product state $({\bf
\Psi}^1_{AB}(t)+{\bf
\Psi}^2_{AB}(t)+{\bf
\Psi}^3_{AB}(t)+{\bf
\Psi}^4_{AB}(t))$.
By linearity:
\begin{displaymath}H_{AB}(t)\cdot {1\over  2}({\bf
\Psi}^1_{AB}(t)+{\bf
\Psi}^2_{AB}(t)+{\bf
\Psi}^3_{AB}(t)+{\bf
\Psi}^4_{AB}(t))=H_{AB}(t)\cdot {1\over 
2}(\big|\psi^1_{A}\big>+\big|\psi^2_{A}\big>)\otimes
(\big|\phi^1_{B}\big>+\big|\phi^2_{B}\big>)\end{displaymath}
\begin{displaymath}={1\over
\sqrt
2}((\big|\triangle_A^{11}\psi^1_{A}\big>+\big|\triangle_A^{21}\psi^2_{A}\big>)\otimes\big|\phi^1_{B}\big>
+(\big|\triangle_A^{12}\psi^1_{A}\big>+\big|\triangle_A^{22}\psi^2_{A}\big>)\otimes\big|\phi^2_{B}\big>\end{displaymath}
\begin{displaymath}
+\big|\psi^1_{A}
\big>\otimes (\big|\triangle_B^{11}\phi^1_{B}\big>+\big|\triangle_B^{12}\phi^2_{B}\big>)+\big|\psi^2_{A}
\big>\otimes (\big|\triangle_B^{21}\phi^1_{B}\big>+\big|\triangle_B^{22}\phi^2_{B}\big>))\end{displaymath}
In virtue of the constraints, we get that:
\begin{displaymath}H_{AB}(t)\cdot {1\over  2}({\bf
\Psi}^1_{AB}(t)+{\bf
\Psi}^2_{AB}(t)+{\bf
\Psi}^3_{AB}(t)+{\bf
\Psi}^4_{AB}(t))=\end{displaymath}
\begin{displaymath}={1\over
\sqrt
2}(\lambda''\big|\psi^1_{A}\big>+\lambda'''\big|\psi^2_{A}\big>)\otimes\big|\phi^1_{B}\big>
+(\big|\triangle_A^{12}\psi^1_{A}\big>+\big|\triangle_A^{22}\psi^2_{A}\big>)\otimes
(\big|\phi^1_{B}\big>+\big|\phi^2_{B}\big>)\end{displaymath}
\begin{displaymath}
+(\big|\psi^1_{A}\big>+\big|\psi^2_{A}\big>)
\otimes (\big|\triangle_B^{11}\phi^1_{B}\big>+\big|\triangle_B^{12}\phi^2_{B}\big>))\end{displaymath}
Such a state does not contain any state bi-orthogonal to ${1\over  2}({\bf
\Psi}^1_{AB}(t)+{\bf
\Psi}^2_{AB}(t)+{\bf
\Psi}^3_{AB}(t)+{\bf
\Psi}^4_{AB}(t))$ only if 
$\lambda''\big|\psi^1_{A}\big>+\lambda'''\big|\psi^2_{A}\big>=\lambda''''(\big|\psi^1_{A}\big>+\big|\psi^2_{A}\big>)$,
which imposes that $\lambda''\,=\,\lambda'''\,=\,\lambda''''$. We can repeat this proof with the indices $ii'$
for the system $A$ and $1j$ for the system $B$ instead of $12$ as it was the case in the previous proof, and
we obtain that
$\big|\triangle_B^{ij}\phi^j_{B}\big>=\big|\triangle_B^{i'j}\phi^j_{B}\big>=\big|\triangle_B^{j}\phi^j_{B}\big>$,
and
$\big|\triangle_A^{ij}\psi^i_{A}\big>=\big|\triangle_A^{i1}\psi^i_{A}\big>-
\lambda(j)\big|\psi^i_{A}\big>=\big|\triangle_A^{i}\psi^i_{A}\big>-\lambda(j)\big|\psi^i_{A}\big>$ (where
$\big|\triangle_A^{i}\psi^i_{A}\big>$ does not depend on $j$ while $\lambda(j)$ and
$\big|\triangle_B^{j}\phi^j_{B}\big>$ do not depend on
$i$). Therefore:
\begin{displaymath}H_{AB}(t)\cdot\big|\psi^i_{A}\big>\otimes
\big|\phi^j_{B}\big>=\big|\triangle_A^{i}\psi^i_{A}\big>\otimes\big|\phi^j_{B}\big>+
\big|\psi^i_{A}\big>
\otimes\big|\triangle_B^{j}\phi^j_{B}\big>-\lambda(j)\big|\psi^i_{A}\big>\otimes\big|\phi^j_{B}\big>\end{displaymath}
which fulfills the eqn. 1 provided we proceed to the following identifications:
$H_A(t)\cdot\big|\psi^i_{A}\big> =
\big|\triangle_A^{i}\psi^i_{A}\big>$ and $H_B(t)\cdot \big|\phi^j_{B}\big> =
\big|\triangle_B^{j}\phi^j_{B}\big>-\lambda(j)\big|\phi^j_{B}\big>$.
This ends the proof of the necessary condition of the main theorem.

B) Proof of the sufficient condition. Let us assume that the full Hamiltonian can be
factorised according to the eqn.\ref{fact}. Let us consider the product state $\psi_{A}^{red}(
t')\otimes\psi_{B}^{red}( t')$, where
$\psi^{red}_{A(B)}( t')$ is a solution of the reduced Schr\"odinger equation
$i
\hbar\,
\partial_{t'}\, 
\psi^{red}_{A(B)}( t') =
 H_{A(B)}( t')\cdot\psi^{red}_{A(B)}( t')$ for the initial condition $\psi^{red}_{A(B)}( t)$
=
$\psi_{A(B)}( t)$. Obviously, $i \hbar\, \partial_{t'}\,
\psi_{A}^{red}( t')\otimes\psi_{B}^{red}( t')=H_{AB}(t')\psi_{A}^{red}( t')\otimes\psi_{B}^{red}(
t')$ and
${\bf
\Psi}_{AB}( t) =
\psi_{A}^{red}( t)\otimes\psi_{B}^{red}( t)$ so that, as the solution of Schr\"odinger is univoquely determined by initial conditions, ${\bf
\Psi}_{AB}( t') =
\psi_{A}^{red}( t')\otimes\psi_{B}^{red}( t')$, $\forall t' \geq t$ which ends the proof of the main
theorem.

{\bf Some remarks and comments:}

The proof of the sufficient condition is also valid for infinitely dimensional Hilbert spaces. We expect that the necessary condition
is also valid in infinitely dimensional Hilbert spaces provided the Hamiltonian is sufficiently
regular, but this is presently a mere conjecture. 

Actually, many results that are presented in the present section already appeared in \cite{Durt} two
 years ago. After the completion of that work, we were kindly informed that very similar results were obtained
independently by J. Gemmer and G. Mahler \cite{10}. In this work, the authors showed that if two quantum
systems mutually interact and that the degree of entanglement remains constant in time for all pure
states (not only factorisable pure states but also entangled ones), the Hamiltonian necessarily
factorises into the sum of individual Hamiltonians. Our main theorem is slightly more general in the
sense that it shows that the same necessary condition can be deduced from the weaker assumption that
all pure factorisable states remain factorisable during their temporal evolution. The authors also
proved that particular pure factorisable states remain factorisable during the evolution if and only
if the eqn.\ref{eff3} is fulfilled, so to say if and only if the the Hamiltonian factorises into
the sum of individual effective Hamiltonians. Although the final results are very close to each
other, both approaches are quite different. For instance, the geometrical properties of the bi-orthogonal
decomposition \cite{5} (see appendix) and of the bi-orthogonality play a crucial role in our proofs while this is not true for what concerns the proofs presented
in \cite{10}  (based on a pseudo-Schr\"odinger-equation) which are more algebrical and less geometrical
than ours. Because of this, our proofs are simpler and more intuitive. The price to pay,
nevertheless, is that our approach remains confined to the situation in which states are
factorisable. Thanks to their more sophisticated mathematical treatment the authors of the ref.\cite{10}
managed to derive an expression aimed at quantifying the amount of entanglement that occurs during
the interaction of two quantum systems that do not remain factorisable. 
Remark that the main theorem and the estimation of the rate of generation of entanglement 
given in appendix are also easy to
prove on the basis of the results obtained by Cirac et al. (refs. \cite{entangling power}) on the entangling power of non-local Hamiltonians, but in the
case of two interacting qubits only.

Finally, it is worth noting that the condition \ref{eff3} encountered in the proof of  the sufficient condition of the
lemma can be generalised to factorisable non-necessarily pure states. This is the essence of the
following theorem that was proven in \cite{Durt} and that we reproduce without proof (the proof is straightforward).

{\bf Theorem 2:}

If initially, a bipartite system is prepared in a factorisable (non-necessarily pure) state: ${\bf
\rho}_{AB}( t=0)=\rho_{A}( t=0)\otimes\rho_{B}( t=0)$, and that $\forall t \geq 0$\begin{equation}
\label{fichtre}H_{AB}( t){\bf
\rho}_{AB}( t)= (H^{eff.}_A( t)\cdot\rho_{A}(
t))\otimes\rho_{B}( t)+\rho_{A}( t)\otimes(H^{eff.}_B( t)\cdot \rho_{B}( t))\end{equation} 
where \begin{displaymath}H^{eff.}_A(t)\cdot\rho_{A}( t)=Tr_B( H_{AB}(
t){\bf
\rho}_{AB}( t))\end{displaymath} and
\begin{displaymath}H^{eff.}_B(t)\cdot
\rho_{B}( t)=Tr_A (H_{AB}( t){\bf \rho}_{AB}( t))-(Tr_{AB} (H_{AB}( t){\bf \rho}_{AB}(
t)))\cdot\rho_{B}( t),\end{displaymath}then, necessarily, the state remains factorisable during the interaction:
${\bf
\rho}_{AB}( t)=\rho_{A}( t)\otimes\rho_{B}( t)$ $\forall t \geq 0$.

In this approach, and with this definition of effective Hamiltonians, we face the following problem:
it is easy to show that the sufficient
condition expressed by the eqn.\ref{fichtre} is also necessary in the case of pure states 
(because then the eqn.
\ref{eff3} must be valid at any time, in virtue of the necessary condition of the lemma but the condition \ref{eff3} 
implies the condition \ref{fichtre} in virtue of the Schr\"odinger equation \ref{schrod},).
 Now, in the case of non-pure states, the sufficient condition
expressed by the eqn.\ref{fichtre} is in general not necessary as
shows the following counterexample. If initially, the bipartite system is prepared in a factorisable
state:
${\bf
\rho}_{AB}( t=0)=\rho_{A}( t=0)\otimes\rho_{B}( t=0)$, and that $\forall t \geq 0, H_{AB}( t)={\bf
\rho}_{AB}( t=0)$, then it is easy to check that ${\bf
\rho}_{AB}( t=0)={\bf
\rho}_{AB}( t) \forall t \geq 0$, $ H^{eff.}_A(t)\cdot\rho_{A}( t)=Tr_B( H_{AB}(
t){\bf
\rho}_{AB}( t))= \rho_{A}^2( t=0)\cdot Tr_{B}\rho^2_{B}( t=0)$, $H^{eff.}_B(t)\cdot
\rho_{B}( t)=Tr_A (H_{AB}( t){\bf \rho}_{AB}( t))-(Tr_{AB} (H_{AB}( t){\bf \rho}_{AB}(
t)))\cdot\rho_{B}( t)=Tr_A\rho_{A}^2( t=0)\cdot \rho^2_{B}( t=0)-Tr_A\rho_{A}^2( t=0)\cdot Tr_{B}\rho^2_{B}(
t=0)\cdot \rho_{B}( t=0)$ and it is easy to check that in general the eqn. \ref{fichtre} is not valid when the initial state is not pure so to say when it is not a product of
pure states. This led us recently to redefine the effective Hamiltonians in order to be able to treat also the case of non-pure states. These results are encapsulated in the following theorem:

 {\bf Theorem 3:}

A bipartite system initially prepared in a factorisable (non-necessarily pure) state (${\bf
\rho}_{AB}( t=0)=\rho_{A}( t=0)\otimes\rho_{B}( t=0)$) remains in a factorisable state throughout the evolution (${\bf
\rho}_{AB}( t)=\rho_{A}( t)\otimes\rho_{B}( t)$ $\forall t \geq 0$)
 if and only if the effect of the Hamiltonian can be factorised as follows:
 $\forall t \geq 0$\begin{equation}
\label{fichtre'}[H_{AB}( t),{\bf
\rho}_{AB}( t)]= ([H^{eff'.}_A( t),\rho_{A}(t)])\otimes\rho_{B}( t)+\rho_{A}( t)\otimes([H^{eff'.}_B( t), \rho_{B}( t)])\end{equation} 
where \begin{displaymath}[H^{eff'.}_A( t),\rho_{A}(t)]=Tr_B( [H_{AB}(
t),{\bf
\rho}_{AB}( t)])\end{displaymath} and
\begin{displaymath}[H^{eff'.}_B(t)\cdot
\rho_{B}( t)]=Tr_A ([H_{AB}( t){\bf \rho}_{AB}( t)])-(Tr_{AB} ([H_{AB}( t),{\bf \rho}_{AB}(
t))])\cdot\rho_{B}( t)=Tr_A ([H_{AB}( t){\bf \rho}_{AB}( t)])\end{displaymath}
 
 It is worth noting that, although their effects are unambiguously defined in terms of the effect of the global Hamiltonian,
  there does not necessarily exist effective Hamiltonians (self-adjoint opertors) $H^{eff'.}$ that satisfy the previous definitions. 
 Therefore the commutators that appear in these definitions must be considered symbolically. 
 Nevertheless, their trace is equal to zero, as would be the case with real commutators.
 
 {\bf Proof of the Theorem 3:}

When we describe the state of the system by a density matrix, its evolution obeys the von Neumann
equation:

\begin{equation}\label{vonneu}i \hbar\, \partial_t\, {\bf \rho}_{AB}( t) =
[ H_{AB}( t),{\bf \rho}_{AB}( t)]\end{equation}where $[X,Y]$ represents the commutator of two operators $X$ and
$Y$.
If the eqn.\ref{fichtre'} is satisfied, we have that:
\begin{displaymath}i \hbar\, \partial_t\, {\bf \rho}_{AB}( t) =
([H^{eff'.}_A( t)\rho_{A}(
t)])\otimes\rho_{B}( t)+\rho_{A}( t)\otimes([H^{eff'.}_B( t),\rho_{B}( t)])\end{displaymath} 

 Let us consider the product state $\rho_{A}^{red}( t)\otimes\rho_{B}^{red}( t)$, where $\rho^{red}_{A(B)}( t)$
is a solution of the reduced von Neumann equation $i \hbar\, \partial_t\, {\bf \rho}^{red}_{A(B)}( t) =
[ H^{eff'.}_{A(B)}( t),{\bf \rho}^{red}_{A(B)}( t)]$ for the initial condition $\rho^{red}_{A(B)}( t=0)$ =
$\rho_{A(B)}( t=0)$. In virtue of the Leibniz rule and of the condition \ref{fichtre'}, we get that $i \hbar\, \partial_t\,
(\rho_{A}^{red}( t)\otimes\rho_{B}^{red}( t))=[H_{AB}(t),\rho_{A}^{red}( t)\otimes\rho_{B}^{red}( t)]$
and ${\bf
\rho}_{AB}( t=0) =
\rho_{A}^{red}( t=0)\otimes\rho_{B}^{red}( t=0)$ so that, as the eqn.\ref{vonneu} is deterministic, ${\bf \rho}_{AB}( t) =
\rho_{A}^{red}( t)\otimes\rho_{B}^{red}( t)$, $\forall t \geq 0$.

Conversely, if ${\bf
\rho}_{AB}( t) =
\rho_{A}^{red}( t)\otimes\rho_{B}^{red}( t) \forall t \geq 0$, then, as the von Neumann evolution \ref{vonneu} is trace preserving, we can, without loss 
of generality, assume that $Tr\rho_{A}^{red}( t)=Tr\rho_{B}^{red}( t)=1,  \forall t \geq 0$ so that, in virtue of the properties of the trace,
 $Tr_{A(B)}( \partial_t\rho_{AB})$ =  $( \partial_t\rho^{red}_{B(A)})$ wich ends the proof. 

 Note that the eqns.(\ref{eff3},\ref{fichtre},\ref{fichtre'}) are linear in the coupling Hamiltonian $H_{AB}$ and are automatically satisfied
when the eqn. \ref{fact} is satisfied. Nevertheless it is non-linear in ${\bf \rho}_{AB}$. Moreover,
the effective potential that acts onto say the $A$ particle is likely to depend on the state of the
$B$ particle, a situation that does not occur if we impose that all product states remain product
states. 

\section{The factorisation approximation and the classical limit.}

{\bf The decoherence program, the PS criterion and the classical limit.}

It is worth noting that the proof of the lemma and
of the theorem 2 as well are also valid when the systems $A$ and $B$ are infinite
dimensional, for instance when they are localised particles that interact through a central potential. In this section
we shall consider only this very simple situation and apply the results of the lemma at the light of the predictability
sieve criterion introduced by
Zurek in the framework of the decoherence approach \cite{Zur1,Zur2,Zur3}.
This program is an attempt to solve fundamental paradoxes of
quantum mechanics (the apparent temporal irreversibility
of the measurement process, the measurement problem that deals with
the separation quantum-classical and so on). 
Two ingredients are essential in this approach:
\begin{itemize}
\item decoherence is seen as an aspect of entanglement (this property is trivial if we measure the 
coherence of a system by the Shannon-von Neumann information of its reduced density matrix, obtained 
after tracing out
the rest of the world (environment)) 
\item the measurement process is seen as the interaction 
between a quantum system (that could even include the supposedly
quantum measuring apparatus) and a complex (supposedly quantum)
environment, typically an infinite bath of oscillators \cite{Zur1,Zur2,Zur3}.
\end{itemize}
Traditionnally, such (open) quantum systems are described by 
semi-phenomenological irreversible equations,
which predict the occurence of an (irreversible) increase
of entanglement between the observed system and the environment.
Decoherence is then directly related to this entanglement increase,
via the well-known properties of Shannon-von Neumann information.
Zurek postulated that, roughly speaking, during the evolution, our brain selected,
during the interactive process of creation of a world view,
the classical islands that correspond to the minimal increase of 
Shannon-von Neumann entropy \cite{Zur2}. This is called the EINselection (EIN for environment induced), and this procedure
has been referred to as the predictability sieve criterion (see the updated version of \cite{Zur1} for a review). 
The emergence of a classical world that obeys EINselections can be explained following two ways: A) they correspond to
maximal (Shannon-von Neumann) information; it is well 
plausible that our brain selects the features of the natural world that contain maximal information; B) we can also invoke an argument of 
structural stability: superposition of states that would belong to such islands would 
be destroyed very quickly by the decoherence process which radiates irremediably the 
coherence (or information or Shannon-von Neumann negentropy) into the environment \cite{Zur1}.
 Up to now, the predictability sieve criterion was
only (to the knowledge of the author) applied to open quantum systems, so to say, it was assumed that 
the environment is complex, with a Poincare recurrence (or revival) time tending to infinity FAPP. 
The decoherence process itself was studied in mesoscopic situations, theoretically and experimentally as well,
 \cite{haroche}, but it was usually assumed that the real, macroscopic situation corresponded to the limit of infinitely small (FAPP) decoherence times. 
In this section, we shall apply the predictability sieve criterion to a very simple situation during which the
system $A$ and the environment $B$ are two distinguishable particles and are described by a (pure) scalar wave function that obeys the non-relativistic
Schr\"odinger equation. We shall also assume that their interaction potential is an action a
distance that is time-independent and invariant under spatial translations (a Coulombian interaction
for instance), this is a standard text-book situation that was deeply studied, for instance in the framework of scattering theory. The systems $A$ and $B$ fulfill thus (in the non-relativistic regime) the following Schr\"odinger
equation:
\begin{displaymath}i \hbar\, \partial_t\, \Psi({\bf r}_{A},\, {\bf r}_{B},\, t) =
-({\hbar^2 \over 2m_A} \Delta_{A}\, +\,  {\hbar^2 \over 2m_B}\Delta_{B})\Psi({\bf r}_{A},\, {\bf
r}_{B},\, t)\end{displaymath}\begin{equation} +\, V_{AB} ({\bf r}_{A}-{\bf r}_{B})\Psi({\bf r}_{A},\, {\bf
r}_{B},\, t) \end{equation}
where $\Delta_{A(B)}$ is the Laplacian operator in the
$A(B)$ coordinates. 

{\bf PS criterion for two interacting particles: the effective field regime.}

Let us now consider that the system $A$ is the quantum system that interests us, and that the other system is its environment. Actually, the argument is symmetrical as we shall see so that this choice is a mere convention.
In order to identify the classical islands, according to the PS criterion, 
we must identify the states that exhibit maximal coherence or maximal Shannon-von Neumann information. We assume here that the full state is pure. Then, the classical islands 
correspond to the states that initially and during their interaction as well,  remain factorisable (more precisely in a pure factorisable state) so that, in virtue of the lemma, 
the Hamiltonian may not couple the state to a biorthogonal state and the eqn.\ref{eff3} is fulfilled. This equation correponds 
to what is somewhat called in the litterature the mean field or effective field approximation. It expresses that everything happens as if 
each particle ($A(B)$) ''felt'' the influence of the other particle as if it was diluted with a probability distribution equal to the 
quantum value $\big| \Psi({\bf r}_{B(A)}\big|^2$. 
It corresponds also to the concept of droplet or diluted particle.
 It can be shown \cite{Durt} that, in the static case, the condition \ref{eff3} reduces to the so-called Hartree approximation \cite{7}.
Let us consider a bound state of the Helium atom for instance, and let us neglect the fermionic exchange
contributions, the spins of the electrons and of the nucleus and so on. The time independent
(electronic) Schr\"odinger equation is then the following:
\begin{equation}E_{AB}\cdot \Psi({\bf r}_{A},\, {\bf r}_{B}) =
(-{\hbar^2 \over 2m_A} \Delta_{A}\, +\,V_A\,-\,  {\hbar^2 \over 2m_B}\Delta_{B}\,+\,V_B)\Psi({\bf r}_{A},\, {\bf
r}_{B})\, +\, V_{AB} ({\bf r}_{A}-{\bf r}_{B})\Psi({\bf r}_{A},\, {\bf
r}_{B}) \end{equation} 
where $V_A$ and $V_B$ represent the external fields (for instance the Coulombian nuclear field), while $V_{AB}$
represents the Coulombian repulsion between the electrons $A$ and $B$. Let us assume that this equation admits a
factorisable solution $\Psi({\bf r}_{A},\, {\bf
r}_{B})$ = $\psi_A({\bf r}_{A})\cdot\psi_B({\bf r}_{B})$; then, as is shown in ref.\cite{Durt}, we can derive, up to elementary manipulations, the following consistency condition:
\begin{displaymath}(E_{AB}-<(-{\hbar^2 \over 2m_A} \Delta_{A}\,
+\,V_A)>_A- <(-{\hbar^2 \over 2m_A} \Delta_{B}\, +\,V_B)>_B  )\cdot \psi_A({\bf r}_{A})\cdot\psi_B({\bf r}_{B})
\end{displaymath}\begin{equation} =( <V_{AB} ({\bf r}_{A}-{\bf r}_{B})>_A + <V_{AB} ({\bf r}_{A}-{\bf
r}_{B})>_B-  V_{AB} ({\bf r}_{A}-{\bf r}_{B})  )\cdot\psi_A({\bf r}_{A})\cdot \psi_B({\bf r}_{B})\end{equation}
Equivalently, when the wave-function does not vanish, the following condition must be satisfied:
\begin{equation}\label{static} V_{AB} ({\bf r}_{A}-{\bf r}_{B})= <V_{AB} ({\bf r}_{A}-{\bf r}_{B})>_A + <V_{AB} ({\bf
r}_{A}-{\bf r}_{B})>_B-  <V_{AB} ({\bf r}_{A}-{\bf r}_{B})>_{AB}  \end{equation}
This is nothing else than the condition \ref{eff3} in a static form. Obviously, in this regime,
particles behave as if they were discernable, and constituted of a dilute, continuous medium
ditributed in space according to the quantum distribution in $\big| \psi_{A(B)}\big|^2(r_{A(B)},t)$.

{\bf Special case 1: the test-particle regime.}

As the potential does depend on the relative position ${\bf r}_{rel}={\bf r}_{A}-{\bf
r}_{B}$ only, it is convenient to pass to the center of mass coordinates:
\begin{displaymath}i \hbar\, \partial_t\, \Psi({\bf r}_{CM},\, {\bf r}_{rel},\, t) =
- ({\hbar^2 \over 2(m_A+m_B)}\Delta_{CM}\, +\,  {\hbar^2 \over 2\mu}\Delta_{rel})\Psi({\bf r}_{CM},\, {\bf
r}_{rel},\, t)\end{displaymath}\begin{equation} +\, V_{AB} ({\bf r}_{rel})\Psi({\bf r}_{CM},\, {\bf
r}_{rel},\, t) \end{equation}where ${\bf r}_{CM}={m_A{\bf r}_{A}+m_B{\bf r}_{B}\over m_A+m_B}$ and
$\mu={m_A\cdot m_B\over m_A+m_B}$. As it is well-known, the previous equation is separable which means that
if, initially, the wave-function is factorisable in these coordinates, it will remain so during the evolution.
Now, we are interested in situations for which the wave-function is initially factorisable according to the
partition of the Hilbert space that is induced by the systems $A$ and $B$. In general, such a wave-function is
not factorisable in the coordinates of the center of mass. Formally, if $\Psi({\bf r}_{A},\, {\bf r}_{B},\, t=0)
= \psi_A({\bf r}_{A},\, t=0)\cdot\psi_B({\bf r}_{B},\, t=0)$, $\Psi({\bf r}_{CM},\, {\bf r}_{rel},\, t=0)=\int
d\omega A(\omega) \psi^{\omega}_{CM}({\bf r}_{CM},\, t=0)\cdot\psi^{\omega}_{rel}({\bf r}_{rel},\, t=0)$ where
$A(\omega)$ is a generally non-peaked amplitude distribution. Then, at time $t$, $\Psi({\bf r}_{CM},\, {\bf
r}_{rel},\, t)=\int d\omega A(\omega)\psi^{\omega}_{CM}({\bf r}_{CM},\, t)\cdot\psi^{\omega}_{rel}({\bf
r}_{rel},\, t),$ where
$\psi^{\omega}_{CM}({\bf r}_{CM},\, t)$ obeyed during the time interval $[0,t]$ a free Schr\"odinger evolution
for the initial condition
$\psi^{\omega}_{CM}({\bf r}_{CM},\, t=0)$ while $\psi^{\omega}_{rel}({\bf r}_{rel},\, t)$ was submitted to the
interaction potential $V_{AB}(r_{rel})$. In general, $\Psi({\bf r}_{A},\, {\bf r}_{B},\, t)$ is no longer
factorisable into a product of the form $\psi_A({\bf r}_{A},\, t)\cdot\psi_B({\bf r}_{B},\, t)$. Actually,
this is not astonishing because, in virtue of Noether's theorem the full momentum is conserved during the
evolution. Therefore the recoil of one of the two particles could be used in order to determine (up to the
initial undeterminacy of the centre of mass) what is the recoil of the second particle. The existence of
such correlations is expressed by the entanglement of the full wave-function. On the basis of such general
considerations we expect that entanglement is very likely to occur due to the interaction between the two
particles. 

Nevertheless, if $m_A<<m_B$, that the initial state is factorisable 
and that the $B$ particle is initially at rest and well localized,
 it can be shown that  the Hamiltonian does not couple the state to a biorthogonal state and the eqn.\ref{eff3} is fulfilled.
  Indeed, if we let coincide the origin of the system of
coordinates associated to the particle
$B$ with its location, and that we neglect its recoil as well as its dispersion, the following approximations are valid: $ {\bf r}_{CM}\approx {\bf
r}_{B}\approx 0$, ${\bf r}_{rel}\approx {\bf r}_{A}-0 ={\bf r}_{A}$, $\psi_A({\bf r}_{A},\,
t)\approx\psi_{rel}({\bf r}_{rel},\, t)$ and $\psi_B({\bf r}_{B},\, t)\approx\psi_{CM}({\bf r}_{CM},\, t)$.
Moreover,
$\Psi({\bf r}_{A},\, {\bf r}_{B},\, t=0) = \psi_A({\bf r}_{A},\, t=0)\cdot\psi_B({\bf r}_{B},\, t=0)\approx
\psi_{rel}({\bf r}_{rel},\, t=0)\cdot\psi_{CM}({\bf r}_{CM},\, t=0)\approx\Psi({\bf r}_{CM},\, {\bf r}_{rel},\,
t=0)$. At time $t$, $\Psi({\bf r}_{CM},\, {\bf r}_{rel},\, t)\approx \psi_{rel}({\bf r}_{rel},\,
t=0)\cdot\psi_{CM}({\bf r}_{CM},\, t)\approx \psi_A({\bf r}_{A},\, t)\cdot\psi_B({\bf r}_{B},\,
t)\approx\Psi({\bf r}_{A},\, {\bf r}_{B},\, t)$. The separability of the full system into its components $A$
and
$B$ is thus ensured, in good approximation, during the scattering process and for bound states as well.

{\bf Special case 2: the material point regime regime.}
Another situation that is of physical interest is the situation of mutual scattering of two well localized 
wave packets  whenever we can neglect the quantum extension of
the interacting particles. This will occur when the interaction potential $V_{AB}$ is smooth enough and that the
particles $A$ and $B$ are described by wave packets the extension of which is small in comparison to the
typical lenght of variation of the potential. It is well known that in this regime, when the de Broglie wave
lenghts of the wave packets are large enough, it is consistent to approximate quantum wave mechanics by
its geometrical limit, which is classical mechanics. For instance the quantum differential cross sections
converge in the limit of short wave-lenghts to the corresponding classical cross section. Ehrenfest's theorem
also predicts that when we can neglect the quantum fluctuations, which is the case here, the average motions are
nearly classical and provide a good approximation to the behaviour of the full wave-packet in so far we
consider it as a material point. In this regime, we can in good approximation replace the interaction potential
by the first order term of its Taylor development around the centers of the wave-packets associated to the
particles $A$ and $B$: \begin{displaymath}V_{AB} ({\bf r}_{A}-{\bf r}_{B})\approx V_{AB} (<{\bf r}_{A}>_t-<{\bf
r}_{B}>_t)+{\bf
\nabla}_A V_{AB}(<{\bf r}_{A}>_t-<{\bf r}_{B}>_t)\cdot({\bf r}_{A}-<{\bf
r}_{A}>_t)\end{displaymath}\begin{displaymath}+{\bf \nabla}_B V_{AB}(<{\bf r}_{A}>-<{\bf r}_{B}>_t)\cdot({\bf
r}_{B}-<{\bf r}_{B}>_t).\end{displaymath} Then the evolution equation is in good approximation separable into
the coordinates
${\bf r}_{A},{\bf r}_{B}$ and we have that, when 
$\Psi({\bf r}_{A},\, {\bf r}_{B},\, t=0) = \psi_A({\bf r}_{A},\, t=0)\cdot\psi_B({\bf r}_{B},\, t=0)$, then, at
time
$t$,
$ \Psi({\bf r}_{A},\, {\bf r}_{B},\, t)\approx\psi_A({\bf r}_{A},\, t)\cdot\psi_B({\bf r}_{B},\, t)$ where 
\begin{displaymath}i \hbar\, \partial_t\, \psi_A({\bf r}_{A},\, t) \approx
-{\hbar^2 \over 2m_A} \Delta_{A}\psi_A({\bf r}_{A},\, t)\end{displaymath}\begin{equation} +\,( V_{AB} (<{\bf
r}_{A}>_t>-<{\bf r}_{B}>_t)+{\bf
\nabla}_A V_{AB}(<{\bf r}_{A}>_t-<{\bf r}_{B}>_t)\cdot({\bf r}_{A}-<{\bf r}_{A}>_t))\psi_A({\bf
r}_{A},\, t) \end{equation}
\begin{displaymath}i \hbar\, \partial_t\, \psi_B({\bf r}_{B},\, t) \approx
-{\hbar^2 \over 2m_B} \Delta_{B}\psi_B({\bf r}_{B},\, t)\end{displaymath}\begin{equation} +\,( {\bf
\nabla}_B V_{AB}(<{\bf r}_{A}>_t-<{\bf r}_{B}>_t)\cdot({\bf r}_{B}-<{\bf r}_{B}>_t))\psi_B({\bf
r}_{B},\, t) \end{equation} Note that the Bohmian velocities associated to the particles $A$ and $B$ are
factorisable only when the full state is factorisable. Otherwise, the velocity of a particle depends non-locally
on the location of both particles.

 In summary, we see thus that, in the simple case considered in this section, the classical islands EINselected in virtue of the PS criterion are regions of the Hilbert space where the 
 mean or effective field approximation (or Hartree approximation in the static case) is valid. Then, the interaction factorises into
the sum of two effective potentials that act separately on both particles, and express the average
influence due to the presence of the other particle (which is not true in general and certainly not
inside the atom). In particular, in the test-particle limit, the effective potential undergone by the
massive particle is close to zero, and when the heavy particle is well localised, its average,
effective, potential is close to the real potential undergone by the light ``test-particle''. In the
classical limit (material points), the quantum internal structure of the interacting particles can be
neglected and the potential is equivalent to the sum of the effective potentials in good approximation. 

Finally, it is worth noting that 
in general the superposition principle is not valid inside classical islands: the dynamical constraints (\ref{eff3},\ref{fichtre},\ref{fichtre'}) being 
non-linear even in the static case (\ref{static}), the superposition of two states that belong to classical islands does not in 
general belong to such an island. Another way to formulate this remark is that decoherence-free subspaces are most often one-dimensional.
 Moreover, 
when the full state is a product of non-pure states, it can happen that, although it remains factorisable in time and although the system is closed 
and undergoes a unitary evolution, which preserves the Shannon-von Neumann entropy of the full state,
 the reduced evolutions do no longer preserve the entropies of the reduced states. 
 In such a situation, the forementioned symmetry between system and environment is broken. This situation is potentially richer because it 
 allows transfers of entropy between the system and the environment but it is out of the scope of the present paper to study 
 all the possibilities that appear in this case.

\section{Conclusions and comments}

Originally, the present work was motivated by the results presented in the references
\cite{8,9}. In these papers it is argued and shown that retrievable, usable quantum information can
be transferred in a scheme which, in striking contrast to the quantum teleportation schemes, requires no external
channel and does not involve the transfer of  a quantum state from one subsystem to the other.
Although other specific quantum ingredients are present (such as entanglement), entanglement-free
interaction between two mutually scattering particles (in the three dimensional, physical space)
plays a crucial role in this scheme. The previous remarks suggest that localisation of at least one
of the particles is one of the necessary ingredients of such protocols for quantum information
transfer. For instance, in the test-particle limit the massive particle is localised while in the
classical limit, both particles are localised. It is easy to show that if at least one of the two
interacting particles is not well localised (bilocated for instance), and that the particles
interact through a position-dependent potential (action at a distance), they are highly likely to
end up in an entangled state.

A conclusion of the first part of this work (main theorem) could be: in quantum mechanics to interact means nearly always to
entangle. We showed that real interactions do necessarily generate entanglement (the inverse result,
that it is impossible to generate entanglement without turning on an interaction, is rather trivial).

Considered
so, the degree of entanglement of large (generic) classes of states for large (generic) classes of systems ought to
increase with time, which would indicate some analogy between entanglement and entropy. Note however
that the temporal reversibility of the Schr\"odinger equation implies that the degree of entanglement
could also decrease in time so that we face a paradox analog to the famous Loschmidt paradox which
emphasises the apparent contradiction between the temporal asymmetry of the second principle of
thermodynamics and the temporal symmetry of fundamental interactions.  Actually, it seems that nowadays the
 Shannon-von Neumann entropy plays more and more the central (and still mysterious) role played by the 
Boltzmann entropy in statistical mechanics and thermodynamics. In quantum information, for instance, it is often implicitly assumed 
that the corresponding negentropy contents the ''reality'' of the system (''all'' is (quantum) information), 
and the decoherence approach confers to (quantum) information, through the PScriterion, 
a supraphysical role: the organisation of our brain (the way we think) as well as the so-called measurement problem 
could be ultimately explained in terms of (quantum) information! Recently, the role played by entanglement during phase-transitions was 
recognised \cite{nature}, and it could be that we shall be able soon to solve old paradoxes such as the Loschmidt paradox in the new framework of quantum information. Obviously, such considerations
are out of the scope of this paper and we invite the interested reader to consult the reference \cite{11}
and references therein.

To the knowledge of the author, nobody else attempted up to know to apply the predictability sieve
 criterion introduced by
Zurek in the very simple and standard situation considered here (the system and the environment as well are described by a pure, scalar wave
 function and their evolution obeys a classical Hamiltonian). 
 It is gratifying to note that according to our analysis presented in the second section all the regimes that
belong to the no-entanglement regime, which is also the
classical regime according to the PS criterion, correspond to classical preconceptions about 
the objects that physicists call particles\footnote{\label{ftn1}Actually, the simple situations considered here also provide right intuitions even when the environment is more complex.
 In refs.\cite{Zur2,Zur3}, Zurek and coworkers considered that the environment consisted of an infinity of oscillators (bath), 
 a very commonly studied case in the theory of open quantum systems.
 In ref.\cite{Zur2}it was assumed that the quantum system itself was an oscillator, and the autors showed
  that the classical islands of the system were the coherent states. 
  It is easy to check that, in our approach,
   if we consider that $V_{A}$ and $V_{B}$ are harmonic,
    and that the oscillators $A$ and $B$ are coupled, at the rotating wave approximation by 
 the standard coupling $V_{AB}=a^+b+ab^+$ where $a$($b$) are the destruction operators for the $A$($B$) quanta, 
 the coherent states obey the eqn. \ref{eff3}. So, the classical (pointer) states for two coupled oscillators
  are the coherent states, which provides the right intuition when one oscillator is coupled to a bath. 
  Similarly, when as in ref.\cite{Zur3} the system is a system with a discrete energy spectrum coupled to a bath of slow oscillators,
   the pointer states (which are in this case the energy eigenstates) are derived modulo the effective field or mean field approximation.} The most elaborated model (the droplet or diluted medium model) appeared 
relatively late in the history (it corresponds to the classical models of the electron developed by Langevin, 
Poincare, Abraham and others at the beginning of our century). In the limit where we can neglect the internal 
structure of the droplet, we recover as special cases the test-particle limit (the internal structure of the massive
particle can be neglected because it is insensitive to the back action of the other one) and the material point regime
(we can neglect the quantum extension of both particles). The concept of test particle emerged in the 19th century, in
the framework of electrodynamics, and was also useful in general relativity. The material point regime corresponds to the
 Ehrenfest quasi-classical equation and also to the geometrical limit of quantum
wave mechanics which is, as is well-known, Hamiltonian mechanics.  The associated concept of material points corresponds
to the Galilean and Newtonian paradigms (particles are like little stones of negligible extent that move in empty
space), a concept that can be traced back to Democritus. It is difficult to find radically different paradigms in
classical physics in order to describe (modellize) the concept of particle. In all the cases, the systems are separated
only in first approximation, and in general entanglement will accompany interaction, in virtue of the main theorem.

One may agree or disagree with our oversimplified choice of what is the system and what is the environment,
 but our analysis shows that even when it is applied in simple situations, the
  PS criterion provides unexpected but very natural and
  useful$^{\ref{ftn1}}$ insights:
 classical islands correspond to the classical preconceptions, and exhibit a certain structural
  stability in time$^{\ref{ftn2}}$.
 It confirms the deep intuition of Schr\"odinger, already mentioned in the
introduction, who described entanglement as the characteristic trait of quantum mechanics, ``the one
that enforces its entire departure from classical lines of thought'' \cite{1}.

\section*{Acknowledgements}
Sincere thanks to John Corbett (Macquarie's University, Sydney) and G. Mahler (Institut fur
Theoretische Physik, Stuttgart) for fruitful discussions and comments. This work originated
during my visit at Macquarie's university in March and April 2001. Support from the Fund for
Scientific Research, Flanders, is acknowledged as well as supports of
the Inter-University Attraction Pole Program of the Belgian government
under grant V-18 and the Concerted Research Action ``Photonics in
Computing and the research council (OZR) of the VUB.

\section*{Appendix. Generation of entanglement during the evolution.}
 Let us consider that at time $t$ the system is
prepared in a product state
${\bf
\Psi}_{AB}( t) =
\psi_{A}( t)\otimes\psi_{B}( t)$, and let us choose a
basis of product states $\big|\psi^i_{A}\big>\otimes \big|\phi^j_{B}\big>$ ($i: 1...d_A;
j:1...d_B$, and $\big<\psi^i_{A}\big|\psi^j_{A}\big> =
\delta_{ij}=\big<\phi^i_{B}\big|\phi^j_{B}\big>)$  such that $\psi_{A}( t) = \big|\psi^1_{A}\big>$
and
$\phi_{B}( t) = \big|\phi^1_{A}\big>$. Then, after a short time $\delta t$, \begin{displaymath}{\bf
\Psi}_{AB}( t+\delta t)= (I+{i\delta t\over \hbar }\cdot H_{AB}(t))\cdot {\bf \Psi}_{AB}( t)+
\tau(\delta t^2)\end{displaymath}where by definition $\tau
(\epsilon^m)$ decreases at least as fast as the $m$th power of $\epsilon$ when $\epsilon$ goes to
zero. In a matricial form, the previous equation becomes:
\begin{equation}{\bf
\Psi}_{AB}( t+\delta t) =\big|\psi^1_{A}\big>\otimes\big|\phi^1_{B}\big>+{i\delta t\over \hbar
}\Sigma_{i:
1...d_A;j:1...d_B}H_{i1j1}\big|\psi^i_{A}\big>\otimes\big|\phi^j_{B}\big>+\tau(\delta
t^2)\end{equation} where
\begin{equation} H_{ikjl}= 
\big<\psi^i_{A}\big|\otimes\big<\phi^j_{B}\big|H_{AB}(t)\big|\psi^k_{A}\big>\otimes\big|\phi^l_{B}\big>+
\tau(\delta t)\end{equation}
Equivalently,
\begin{displaymath}{\bf \Psi}_{AB}(
t+\delta t)= \big|\psi^1_{A}\big>\otimes\big|\phi^1_{B}\big>+{i\delta t\over \hbar}(\Sigma_{i:
1...d_A}H_{i111}\big|\psi^i_{A}\big>\otimes\big|\phi^1_{B}\big>\end{displaymath}
\begin{displaymath}+\Sigma_{j:
2...d_B}H_{11j1}\big|\psi^1_{A}\big>\otimes\big|\phi^j_{B}\big>+\Sigma_{i:
2...d_A;j:2...d_B}H_{i1j1}\big|\psi^i_{A}\big>\otimes\big|\phi^j_{B}\big> )+ \tau(\delta
t^2)\end{displaymath}All the components of ${\bf \Psi}_{AB}(
t+\delta t)$ that are bi-orthogonal to ${\bf
\Psi}_{AB}( t)$ are contained in the last term of
the previous equation: $\Sigma_{i:
2...d_A;j:2...d_B}H_{i1j1}\big|\psi^i_{A}\big>\otimes\big|\phi^j_{B}\big> )$, up to $\tau(\delta
t^2)$. We can
rewrite this equation as follows:
\begin{displaymath}{\bf \Psi}_{AB}(
t+\delta t)= (\big|\psi^1_{A}\big>+{i\delta t\over \hbar
}\Sigma_{i:
1...d_A}H_{i111}\big|\psi^i_{A}\big>)\otimes(\big|\phi^1_{B}\big>+{i\delta t\over \hbar
}\Sigma_{j:
2...d_B}H_{11j1}\big|\phi^j_{B}\big>)\end{displaymath}
\begin{equation}\label{equation}+{i\delta t\over \hbar
}\Sigma_{i:
2...d_A;j:2...d_B}H_{i1j1}\big|\psi^i_{A}\big>\otimes\big|\phi^j_{B}\big>+ \tau(\delta
t^2)\end{equation}
In virtue of the necessary condition of the lemma, if the Hamiltonian couples ${\bf \Psi}_{AB}(
t=0)$ to states that are bi-orthogonal to it, which means that $\Sigma_{i:
2...d_A;j:2...d_B}\big|H_{i1j1}\big|^2\not= 0,$ then, the full state does not remain factorisable. Actually, we have
that the
development of the first order in $\delta t$ of the bi-orthogonal or Schmidt decomposition \cite{5} of ${\bf \Psi}_{AB}(
t+\delta t)$ contains more than one product state, which means that ${\bf \Psi}_{AB}(
t+\delta t)$ is entangled for $\delta t$ small enough. In order to prove it, let us consider the
components of
${\bf
\Psi}_{AB}( t+\delta t)$ that are bi-orthogonal to ${\bf
\Psi}_{AB}( t)$. In virtue of Schmidt's theorem of the bi-orthogonal decomposition \cite{5}, we can
find $d_A-1$ normalized states
$\big|\tilde
\psi^i_{A}\big>$ of
${\bf C}^{d_A}$ mutually orthogonal and orthogonal to $\big|
\psi^1_{A}\big>$ and $d_B-1$ normalized states $\big|\tilde \phi^j_{B}\big>$  of ${\bf
C}^{d_B}$ mutually orthogonal and orthogonal to $\big|
\phi^1_{B}\big>$ such that  \begin{displaymath}{i\delta t\over \hbar}\Sigma_{i:
2...d_A;j:2...d_B}H_{i1j1}\big|\psi^i_{A}\big>\otimes\big|\phi^j_{B}\big>=\Sigma_{i:
2...min(d_A,d_B)}\alpha_i \big|\tilde \psi^i_{A}\big>\otimes \big|\tilde
\phi^i_{B}\big>\end{displaymath}
Let us now define the state $\big|\tilde \psi^{1'}_{A}\big>$ of ${\bf C}^{d_A}$ as follows:
$\big|\tilde \psi^{1'}_{A}\big>={1\over N_1}\cdot (\big|\psi^1_{A}\big>+{i\delta t\over \hbar
}\Sigma_{i:
1...d_A}H_{i111}\big|\psi^i_{A}\big>)$, where $N_1$ is a normalisation factor, and let us replace
the orthonormal basis $\{\big| \psi^1_{A}\big>,\big|\tilde \psi^2_{A}\big>,\big|\tilde
\psi^3_{A}\big>,...,\big|\tilde \psi^{d_A}_{A}\big> \}$ of ${\bf C}^{d_A}$ by the orthonormal
basis $\{\big| \tilde \psi^{1'}_{A}\big>,\big|\tilde \psi^{2'}_{A}\big>,\big|\tilde
\psi^{3'}_{A}\big>,...,\big|\tilde \psi^{d_A'}_{A}\big> \}$ of ${\bf C}^{d_A}$ that we
obtain by the Gram-Schmidt orthonormalisation procedure:
\begin{displaymath}\big|\tilde \psi^{2'}_{A}\big>={1\over N_2}\cdot
(\big|\tilde \psi^2_{A}\big>- \big< \tilde \psi^{1'}_{A}\big|\tilde \psi^2_{A}\big>\cdot  \big|
\tilde
\psi^{1'}_{A}\big>)\end{displaymath}where $N_2$ is a normalisation factor.
\begin{displaymath}\big|\tilde \psi^{3'}_{A}\big>={1\over N_3}\cdot
(\big|\tilde \psi^3_{A}\big>- \big< \tilde \psi^{1'}_{A}\big|\tilde \psi^3_{A}\big>\cdot  \big|
\tilde
\psi^{1'}_{A}\big>- \big< \tilde \psi^{2'}_{A}\big|\tilde \psi^3_{A}\big>\cdot  \big|
\tilde
\psi^{2'}_{A}\big>)\end{displaymath}where $N_3$ is a normalisation factor, and so on. It is easy to
check that $\big|
\tilde
\psi^{i'}_{A}\big>=\big|
\tilde
\psi^i_{A}\big>+\tau(\delta t).$ Note that this is no longer true when the dimension $d_A$ is not
finite. We can repeat the same operation in order to replace
the orthonormal basis $\{\big| \psi^1_{B}\big>,\big|\tilde \psi^2_{B}\big>,\big|\tilde
\psi^3_{B}\big>,...,\big|\tilde \psi^{d_B}_{B}\big> \}$ of ${\bf C}^{d_B}$ by the orthonormal
basis $\{\big| \tilde \psi^{1'}_{B}\big>,\big|\tilde \psi^{2'}_{B}\big>,\big|\tilde
\psi^{3'}_{B}\big>,...,\big|\tilde \psi^{d_B'}_{b}\big> \}$ of ${\bf C}^{d_B}$. Then,
after substitition in the eqn. \ref{equation}, we obtain that: 
\begin{displaymath}{\bf \Psi}_{AB}(
t+\delta t)= \Sigma_{i:
1...min(d_A,d_B)}\alpha_i \big|\tilde \psi^{i'}_{A}\big>\otimes \big|\tilde
\phi^{i'}_{B}\big>+\tau(\delta t^2)\end{displaymath}where
\begin{equation}\label{schmidt}\big|\alpha_1\big|^2=1+\tau(\delta t^2), \Sigma_{i:
2...min(d_A,d_B)}\big|\alpha_i\big|^2 = {\delta t^2\over \hbar^2}\Sigma_{i:
2...d_A;j:2...d_B}\big|H_{i1j1}\big|^2+\tau(\delta t^3)\end{equation} The previous equation
expresses that the development up to the first order in $\delta t$ of the bi-orthogonal Schmidt
decomposition of ${\bf \Psi}_{AB}( t+\delta t)$ contains more than one product state. It is well
known that then ${\bf \Psi}_{AB}( t+\delta t)$ is an entangled state. Nevertheless, for those who
are not familiar with this property, we shall prove directly the result by estimating the linear entropy of the
reduced density matrix. By definition, the reduced density matrix $\rho_A$ of the system $A$ is
equal to $Tr_B \rho$ where
$\rho$ is the projector on ${\bf \Psi}_{AB}$. Obviously, when the state of the system is a product
state (${\bf \Psi}_{AB} =\psi_{A}\otimes\psi_{B}$), $\rho_A$ is the projector on $\psi_{A}$, and
we have that $\rho_A$ = $\rho_A^2$, and $Tr\rho_A$ = $Tr\rho_A^2$ = 1. 

As we mentioned before, $Tr\rho_A$ -
$Tr\rho_A^2$ provides a good measure of the degree of the entanglement of the full system, in the vicinity of product
states.

If the Schmidt bi-orthogonal decomposition of the state ${\bf \Psi}_{AB}$ is equal to $\Sigma_{i:
1...min(d_A,d_B)}\alpha'_i \big| \psi^{i'}_{A}\big>\otimes \big|
\phi^{i'}_{B}\big>$, then it is easy to check that $\rho_A = \Sigma_{i:
1...min(d_A,d_B)}\big|\alpha'_i \big|^2\big| \psi^{i'}_{A}\big>\big< \psi^{i'}_{A}\big|$, $Tr\rho_A =
\Sigma_{i: 1...min(d_A,d_B)}\big|\alpha'_i \big|^2=1$ by normalisation and $Tr\rho_A^2 =
\Sigma_{i: 1...min(d_A,d_B)}\big|\alpha'_i \big|^4\leq (Tr\rho_A)^2 = 1^2=1.$ The last inequality
is saturated for product states only. Note that $Tr\rho_A^2= Tr\rho_B^2$ which shows that this
parameter expresses properties of the system considered as a whole, as it must be when we are dealing
with entanglement. Obviously
$Tr\rho_A^2(t+\delta t)=
\big|\alpha_1
\big|^4+ \Sigma_{i:
2...min(d_A,d_B)}\big|\alpha_i \big|^4$ and $\Sigma_{i:
2...min(d_A,d_B)}\big|\alpha_i \big|^4\leq (\Sigma_{i:
2...min(d_A,d_B)}\big|\alpha_i \big|^2)^2$

But $\big|\alpha_1
\big|^4 = (1-\Sigma_{i:
2...min(d_A,d_B)}\big|\alpha_i\big|^2)^2 = (1-{\delta t^2\over \hbar^2}\Sigma_{i:
2...d_A;j:2...d_B}\big|H_{i1j1}\big|^2+\tau(\delta t^3))^2$ and $(\Sigma_{i:
2...min(d_A,d_B)}\big|\alpha_i \big|^2)^2 = ({\delta t^2\over \hbar^2}\Sigma_{i:
2...d_A;j:2...d_B}\big|H_{i1j1}\big|^2+\tau(\delta t^3))^2=\tau(\delta t^4)$ in virtue of the eqn. \ref{schmidt}
so that $Tr\rho_A^2(t+\delta t)= 1-2 \cdot {\delta t^2\over \hbar^2}\Sigma_{i:
2...d_A;j:2...d_B}\big|H_{i1j1}\big|^2+\tau(\delta t^3)$ for $\delta t$ small
enough, which proves that, for time increments short enough, the (quadratic in time) squared norm of the
biorthogonal component
that is generated during the evolution is equal to one halve of the increase of the  linear entropy production.  As we
noted, the first temporal derivative of this increase is always equal to zero for product states. This reflects an
interesting geometrical property: in
comparison to
 other states, product states are global minima of the linear entropy (this is also true for the Shannon-von Neumann
entropy) \footnote{\label{ftn2}It can be shown by direct computation that when the state of the system is a
product state (${\bf \Psi}_{AB}(t) =\psi_{A}(t)\otimes\psi_{B}(t)$), then the following
identity ${dTr\rho_A^2\over dt}(t)=0$ is necessarily satisfied, independently of the form of the
Hamiltonian $H_{AB}$. This explains why no term of the first order in time appears in the
previous development. This is a good point in the favour of the PS criterion where it is assumed that such states exhibit a 
certain structural stability.}.

\begin{thebibliography}{99}



\bibitem{1} E. Schr\"odinger, {\it Discussion of probability relations between separated systems}, Proc.
Cambridge Philos. Soc. {\bf 31}, 555 (1935).

\bibitem{2} J. S. Bell, {\it On the EPR paradox}, Physics, {\bf 1}, 165 (1964).


\bibitem{3} N. Gisin, {\it Bell's inequality holds for all non-product states}, Phys. Lett. A {\bf 154}, n$^o$
5,6, 201 (1991).

\bibitem{4} D. Home and F. Selleri, {\it Bell's
theorem and the EPR paradox}, La Rivista del Nuovo Cimento della Societa Italiana di fisica, {\bf 14}, n$^o$ 9
(1991) p 24. 

\bibitem{Zur1} W.H. Zurek, {\it Decoherence and the transition from quantum to classical},
 Physics Today, 
{\bf 10},  36, (Oct. 1991) (see also the updated version: quant-ph/0306072).

\bibitem{Zur2} W.H. Zurek, S. Habib, and J.P. Paz {\it Coherent states via decoherence},
 Phys. Rev. Lett.,
{\bf 70}, n$^o$9, 1187, (1993).

\bibitem{Zur3}  J.P. Paz and W.H. Zurek,  {\it Quantum limit of decoherence: environment induced 
superselection of energy eigenstates},
 Phys. Rev. Lett., {\bf 82}, n$^o$26, 5181, (1999).


\bibitem{Durt} T. Durt, {\it Characterisation of an entanglement-free evolution}, quant-ph/0109112
(2001).

\bibitem{10}J. Gemmer and G. Mahler, {\it Entanglement and the Factorization-Approximation},
quant-ph/0109140, Eur. Phys J. D 17, 385 (2001).

 \bibitem{5} A. Peres, {\it Quantum Theory: Concepts and Methods}, Kluwer Dordrecht (1993) p123.

\bibitem{entangling power} B. Kraus and J.I. Cirac, {\it Optimal creation of entanglement capabilities 
 using a two-qubit gate}, Phys. Rev. A 63, 062309 (2001),
   W. D\"ur, G. Vidal, J.I. Cirac, N. Linden and S. Popescu, 
{\it Entanglement capabilities of non-local Hamiltonians},
 Phys. Rev. Lett. {\bf 87}, 137901 (2001), J.I. Cirac, W. D\"ur, B. Kraus and M. Lewenstein: 
{\it Entangling operations and their implementation using a small amount of entanglement},
 Phys. Rev. Lett., {\bf 86}, n$^o$3, 544, (2001).
 
 \bibitem{haroche}J.M. Raimond, M. Brune and S. Haroche, 
{\it Reversible decoherence of a mesoscopic superposition of field states},
 Physical Rev. Lett. {\bf 79},  n$^o$11, 1964 (1997).




\bibitem{7} L. Landau and E. M. Lifshitz, {\it Non-Relativistic Quantum Mechanics}, Pergamon Press
Oxford (1962) p234.


\bibitem{8} J. Corbett and D. Home, {\it Quantum effects involving interplay between unitary
dynamics and kinematic entanglement}, Phys. Rev. A, {\bf 62}, 062103 (2000).

\bibitem{9} J. Corbett and D. Home, {\it Ipso-Information-transfer}, quant-ph/0103146.

Ê


\bibitem{nature}A. Osterloh, L. Amico, G. Falci and R. Fazio, 
{\it Scaling of entanglement close to a quantum phase transition}, Letters to Nature, {\bf 416}, 608 (2001).

\bibitem{11} J. Gemmer, A. Otte and G. Mahler, {\it Quantum approach to a derivation of the second
law of thermodynamics}, Phys. Rev. Lett. {\bf 86}, n$^o$10, 1927 (2001).





\end{thebibliography}
\end{document}